\documentclass{article}

\usepackage[nonatbib,final]{nips_2016}

\usepackage[utf8]{inputenc} % allow utf-8 input
\usepackage[T1]{fontenc}    % use 8-bit T1 fonts
\usepackage{hyperref}       % hyperlinks
\usepackage{url}            % simple URL typesetting
\usepackage{booktabs}       % professional-quality tables
\usepackage{amsfonts}       % blackboard math symbols
\usepackage{nicefrac}       % compact symbols for 1/2, etc.
\usepackage{microtype}      % microtypography
\usepackage{graphicx,wrapfig}
\usepackage{float}
\usepackage{amsmath}
\title{Resolving Implicit Coordination in Multi-Agent Deep RL with Deep Q-Networks \& Game Theory}

\author{
    Griffin Adams \\
   \texttt{gta@cs.cmu.edu}
  \And
  Sarguna Janani Padmanabhan \\
 \texttt{sjpadman@andrew.cmu.edu} 
  \And
  Shivang Shekhar \\
  \texttt{shivangs@andrew.cmu.edu} 
}

\begin{document}
\maketitle

\begin{abstract}

We address two major challenges of implicit coordination in multi-agent deep reinforcement learning: non-stationarity and exponential growth of state-action space, by combining Deep-Q Networks for policy learning with Nash equilibrium for action selection. Q-values proxy as payoffs in Nash settings, and mutual best responses define joint action selection. Coordination is implicit because multiple/no Nash equilibria are resolved deterministically. We demonstrate that knowledge of game type leads to an assumption of mirrored best responses and faster convergence than Nash-Q. Specifically, the Friend-or-Foe algorithm demonstrates signs of convergence to a Set Controller which jointly chooses actions for two agents. This encouraging given the highly unstable nature of decentralized coordination over joint actions. Inspired by the dueling network architecture \cite{wang2015dueling}, which decouples the Q-function into state and advantage streams, as well as residual networks \cite{he2016deep}, we learn both a single and joint agent representation, and merge them via element-wise addition. This simplifies coordination by recasting it is as learning a residual function.  We also draw high level comparative insights on key MADRL and game theoretic variables: competitive vs. cooperative, asynchronous vs. parallel learning, greedy versus socially optimal Nash equilibria tie breaking, and strategies for the no Nash equilibrium case. We evaluate on 3 custom environments written in Python using OpenAI Gym \cite{brockman2016openai}: a Predator Prey environment, an alternating Warehouse environment, and a Synchronization environment. Each environment requires successively more coordination to achieve positive rewards.  All code is available on this \href{https://github.com/psjanani/RLAProject}{GitHub link}.

\end{abstract}

\section{Introduction} \label{intro}
A societal shift towards autonomous systems necessitates efficient, scalable, and fault-tolerant coordination among autonomous agents. While centralized command is most effective, network costs and exposure to a single source of failure favor decentralized coordination.  In multi-agent deep reinforcement learning (MADRL), optimal control depends not only on the environment but on the policies of other agents. Effective control requires either implicit coordination or direct competition.  Roadblocks to optimal outcomes include non-stationarity and state-action sparsity \cite{busoniu2008comprehensive}.  Non-stationarity encapsulates the `moving-target' problem, in which agents must continuously co-adapt to each other’s behavior.  Agents are forced to negotiate a trade-off between learning stability and behavior adaptation \cite{busoniu2008comprehensive}. Most MADRL algorithms require parameterizing the joint state-action space, an exponentially increasing space which with respect to each additional agent. A sparsity problem exists because the majority of state-action pairs are never visited.  We address both issues: sparsity and non-stationarity, via an exhaustive search through the joint space of deep architectures (for state value estimation) and game theory (for decentralized policy control).  Game theory enables implicit coordination through unilaterally enforced rules with minimal preconditions.

\subsection{Related Work}
Developments in deep reinforcement learning have resulted in super-human performance on simulated Atari games \cite{mnih2015human}. Mnih et al. \cite{mnih2015human} introduce a novel variant on Q-Learning which estimates q-values using a deep convolutional architecture. The success of the DQN framework has encouraged many extensions to the deep network architecture with the goal of addressing state-action sparsity and non-stationarity. Wang et al. \cite{wang2015dueling} tackle the sparsity issue by decomposing the Q-function into a state value function and a state-dependent function. Their novel dueling-network architecture enables agents to distinguish between non-valuable and valuable states, in which some actions maintain a relative advantage over others. The authors obtain state of the art performance on Atari games. We implement dueling Q-networks for our multi-agent models which directly parametrize the joint action space (referred to in Section \ref{dqns} as Joint-DQN models). Extensions, such as double Q-networks \cite{van2016deep}, minimize q-value over-estimation and maximize learning stability through use of a target network.

Game theoretic approaches to MADRL offer rule-based approaches to deterministically solving the problem of decentralized coordination. The research focuses on policy control and softens the over-reliance of most other literature on learning q-values.  Q-values proxy as payoffs and Nash equilibria are calculated over the joint action space \cite{hu2003nash}. In Nash-Q learning, agents keep track of their own Q-function as well as other agents', and compute temporal difference (TD) targets with the assumption each agent will play its own Nash strategy \cite{hu2003nash}. General-sum stochastic games can have multiple Nash equilibria which has traditionally made the application of game theory to RL a ``nondeterministic procedure'' \cite{{shoham2003multi}}. Yet in the presence of tie breaking rules, Nash-Q learning and its variants \textit{can} converge to a stationary policy under restrictive and debated conditions. Shoham et al. \cite{shoham2003multi} argue a centralized tie-breaking "oracle" is needed (what we deem in our experiments, a Set Controller), while Hu et al. \cite{hu2003nash} argue that Nash-Q learning will converge in the presence of a globally optimal joint action profile at each step. There is no consensus opinion, however, and convergence with Nash-Q appears to be very sensitive to game dynamics. (As such, we test on 3 environments). The Friend-or-Foe (FoF) algorithm represents a simple extension of Nash-Q: an agent designates others as friends or foes and presumes the other agents either seek to minimize (zero-sum foes), or maximize (abetting friends), its q-values \cite{shoham2003multi,littman2001friend}. FoF's assumption of a globally optimal joint action profile enables action selection to occur over one's own q-values without consideration of the other agent's q-values.  Inspired by the correlated equilibrium concept, Greenwald et al. introduce correlated Q-learning (CQ-Learning) \cite{greenwald2003correlated}, which deliberates among equilibria (i.e., Nash) according to several rules: maximizing the sum of agents' q-values (Utilitarian), maximizing the minimum of agents' q-value (Egalitarian), and maximizing the maximum q-value (Republican).

\section{Methodology}

Our methods leverage the success of the DQN architecture \cite{mnih2013playing} and extend it to a niche class of general-sum common-payoff games.  These games feature multiple agents who share rewards across multi-step finite episodes.  We restrict ourselves to games which can be modeled as Markov Decision Processes (MDP) with discrete state and action spaces. We only consider model free learning where the agent has full knowledge of the current state yet must formulate an experimentally driven estimate of transition probabilities.

\textbf{Independent-DQN versus Joint-DQN}\label{dqns}: We explore the two main challenges of MADRL: non-stationarity and state-action sparsity \ref{intro}, through a experimental grid search of architectures and learning procedures.  The primary theoretical decision relates to whether agents learn the full joint action space, or learn \textit{independently} of the other agents. We have created two agent phenotypes: \textbf{Independent-DQN} (IDQN) and \textbf{Joint-DQN} (JDQN). IDQN extends the DQN algorithm directly to the multi-agent setting by treating other agents as part of the environment.  For notation purposes, consider the transition tuple $(\phi_j, a_j, r_j, \phi_{j+1})$: , where $\phi_j$ and $\phi_{j+1}$ represent the current and subsequent processed state frames, $a_j$ represents the action chosen at time $j$, and $r_j$ the reward at time $j$. Each IDQN $agent_i$ models its own TD-target for transition $j$ as
$$ r_j + \gamma \max_{a'} Q(\phi_{j+1}, a'; \theta_i).$$
$Q$ represents a deep network with $\theta_i$ parameters. $\gamma$ represents the reward discount factor. The gradient update is performed on mini-batches according to the Huber loss \cite{huber1964robust} of the TD target against the predicted q-value $Q(\phi_j, a_j; \theta)$. The JDQN agent learns the joint action space generated by all agents. Formally speaking, $agent_i$ models its TD-target for transition tuple $j$ as
$$r_j + \gamma Q(\phi_{j+1}, Joint(Q(\phi_{j+1},\vec{a},1; \theta_1 ),Q(\phi_{j+1},\vec{a},i; \theta_i ), ... ,Q(\phi_{j+1},\vec{a},N; \theta_N )); \theta_i)$$
where $\vec{a}$ represents the joint action vector, $N$ represents the number of agents, and $\theta_i$ represents the network parameters for the $i^{th}$ agent.  The agent’s unique id is passed in to the Q network since it must first transform a global state representation into its own frame of reference. The function $Joint$ stands for any function which receives q-values for each agent, and produces a joint action vector. Above, we considered the scenario where each agent has access to the q-network parameters of the other agents. In the case where the agent has no record, either self-maintained or requested, of the other agents' network parameters, it uses its own as a proxy. The TD target for the hidden q-value case for $agent_i$ is
$$r_j + \gamma Q(\phi_{j+1}, Joint(Q(\phi_{j+1},\vec{a},1; \theta_i ),Q(\phi_{j+1},\vec{a},i; \theta_i ), ... ,Q(\phi_{j+1},\vec{a},N; \theta_i )); \theta_i)$$
which relies solely on a single set of network parameters $\theta_i$. We also consider a further simplification of the Nash-Q: The Friend-or-Foe algorithm \cite{littman2001friend}. The Friend (cooperative case) algorithm assumes a globally optimal joint action and hence greedily selects actions according to just its own q-values. It fully expects the other agent will greedily select the mirror image joint action profile. In theory, fully cooperative settings should have mirrored best responses which makes greedy action over the joint action space synonymous with computing a Nash equilibrium. Yet, Q-functions trained over separately generated data will not always (initially) exhibit mirrored best responses. Hence, the Friend algorithm represents a simplification to which the private Nash-Q setting should approach. Yet the Friend algorithm acts independently over the joint space and hence maintains a stabler control function. In this light, the Friend reduction is both theoretically sound and more efficient.

\textbf{Policy Control}: For training purposes, policy control is carried out according to separate $\epsilon$-greedy linear decay policies for the IDQN architecture.  For the JDQN architecture, action vectors are selected according to a $\epsilon$-joint policy with linear decay. With probability $\epsilon$, random actions are selected, and with probability $1 - \epsilon$, the $Joint$ function chooses the action vector.

\textbf{Asynchronous Single Network versus Parallel}\label{async-explanation}: To probe the non-stationarity issue, we consider two methods for maintaining DQNs. The first method, Asynchronous Single, uses a single Q-network and buffer. All agents’ experience contributes to the buffer which provides sample transitions to update the network. Each agent maintains a copy of the network and asynchronously requests its parameters. The second approach, Parallel, maintains a network and buffer for each agent. Learning occurs in parallel without any experience or weight sharing.

\textbf{Single Stream versus Split Stream}: Inspired by the success of residual networks, we explore recasting agent interaction as learning a residual function above and beyond the single agent state value estimation.  Concretely, we learn separate upscaled single agent and other agent representations. We then concatenate the representations and learn a joint embedding, which is then added element-wise to the single agent's upscaled representation. We model the archetypal residual building block:

\begin{figure}[H]
    \centering
    \includegraphics[width=60mm, height=30mm]{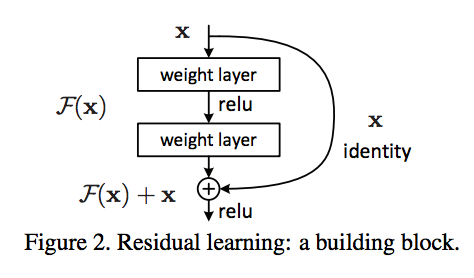}
    \caption{A skip connection building block from the seminal paper \cite{he2016deep}}
\end{figure}

Where, in our case, $F(x)$ represents the joint embedding of the single agent state information with that of the other agent's, and $x$ the learned single agent state representation.

\textbf{Competitive versus Cooperative}: We consider competitive rewards, where rewards are not shared, and cooperative, where agents receive a common payoff regardless of the triggering agent.

\textbf{Game theory \& Implicit Coordination}\label{gametheory}: \label{nash-explanation} For the JDQN architecture, we consider several game theoretic approaches for joint action selection. Inspired by Nash-Q learning \cite{busoniu2008comprehensive,hu2003nash}, we treat q-values as payoffs and compute a payoff matrix for each agent.  In the two player case with two actions, the payoff matrix for state $S$ is as follows:
 
\begin{centering}
\textbf{Agent 2}
$$
\begin{array}{cc}
\textbf{Agent 1} & \begin {array}{cc}
    Q(S, [a_1^1,a_1^2]; \theta_1 ),  Q(S, [a_1^1, a_1^2]; \theta_2) & Q(S, [a_1^1, a_2^2]; \theta_1), Q(S, [a_1^1, a_)^2]; \theta_2)  \\
    Q(S, [a_2^1, a_1^2]; \theta_1), Q(S, [a_2^1, a_1^2]; \theta_2)       & Q(S, [a_2^1, a_2^2]; \theta_1),  Q(S, [a_2^1, a_2^2]; \theta_2)  \\
\end{array}

\end{array}$$
\end{centering}

\noindent where $\theta_i$ represents agent $i$’s Q-network parameters, $N$ represents the number of agents, and $a_i^j$ represents the $i$’th action played by agent $j$.  The best response actions for agent 1, then, represent the row-wise maximum over the first terms of each payoff matrix cell.  The best response actions for agent 2 represent the column-wise maximum over the second terms of the payoff matrix cells.  Nash equilibria are stable in the sense that no agent has an incentive to unilaterally deviate.  Importantly, however, it does not necessarily represent the social optimum. Additionally, rules must exist for the case of multiple or no Nash equilibria. We consider two approaches for the multiple equilibria outcome. For \textit{greedy} selection, each agent acts according to the Nash equilibria which maximizes its own q-value (\textbf{Greedy-Tie-Break}). This may or may not result in a true Nash equilibrium since agents may decide on actions from different equilibria. As the gold standard case, we consider a socially optimal scheme, where each agent acts according to the Nash equilibria with the greatest q-value sum (\textbf{Max-Sum-Tie-Break}). We consider control rules for the no Nash equilibrium scenario. One rule, \textbf{Greedy-No-Nash}, stipulates agents greedily take the action corresponding to their highest respective q-values. Another approach, \textbf{Best-Sum-No-Nash}, stipulates each agent treat the other agents’ actions as equiprobable and choose the action corresponding to the highest average q-value.

\section{Experimental Setup}

\subsection{Environments}

We have built three custom environments using OpenAI gym’s \cite{brockman2016openai} python API: Predator Prey, Warehouse, and Sync. They require increasingly more advanced levels of coordination. To conserve space, we show results only for the latter 2, more advanced environments.

\subsubsection{Predator Prey} 

A multi-player game of tag: predators chase prey and are rewarded when occupying the same cell as a prey. The terminal state is reached when all prey are captured. Prey are considered part of the environment and predators are the trainable agents. Predators and prey can move in any of four directions one cell at a time. Boards are represented as 2-D arrays.

\subsubsection{Warehouse}\label{warehouse-env} 

A simulated warehouse. The warehouse is a 7x7 grid similar to the Predator Prey state representation.  It has two shelves which exist in opposite corners of the board. An episode starts with a single box placed in one of the corners.  A single agent must deliver the box to the center of the board in which case a positive reward is given, and a new box is placed on the shelf \textit{opposite} the previous box placement. This encourages one agent to pick up and drop off a box and the other agent to wait at the other shelf, for which a box will be deposited as soon as the other agent drops off the previous box. The episode ends when $n$ boxes have been deposited. Actions involve moving up, right, down, or left.

\subsubsection{Sync Environment}\label{sync-env} A toy synchronization game. Two agents are situated randomly on a $5x5$ grid and most arrive at opposite corners of the board in the \textit{same} time step. The episode reaches terminal state when one or both of the agents reach one of the two corners (upper left or bottom right).  Agents only receive positive reward if the episode ends with both corners covered. This necessitates considering the likely action of the other agent. Actions involve moving up, right, down, left, or stationary.

\subsection{State Representation}

All experiments are run on 2 agent variants of the above environments. Each agent receives its own location as a row-column tuple, as well as the location of the other agent.  For the Warehouse environment, an additional per-agent marker relates to whether or not that agent is currently carrying a box. For the warehouse environment, since the box location is stochastic, the row and column of a box to be picked up is also provided.  The state representation for the Warehouse Environment is:

$$ [ row_1, col_1, is\_carrying_1, row_2, col_2, is\_carrying_2, row\_box, col\_box ] $$

Where subscripts denote the respective agent ids and $[row\_box, col\_box]$ the grid location of the box. For the Sync Environment the state representation is simply $ [ row_1, col_1, row_2, col_2 ] $.

\subsection{Network Architectures}

We train and test on two deep linear architectures.

\textbf{Single Stream}: The single stream DQN architecture is a three-layer MLP with sigmoid activations.  The output layer is of like-dimension to the number of actions.  For J-DQN settings, the joint action space is of dimension $16$ for the 2-player Warehouse and Predator Prey environments and $25$ for the 2-player Sync environment (extra stationary action).

\textbf{Split Stream}: \label{split-stream-explanation} The split stream architecture decouples the state representation into self location and other agent location.  The locations are separately passed through a single layer MLP with sigmoid activation to for upscaling.  Then, the upscaled representations are concatenated and passed through a dense layer with ReLu activation. This layer learns a joint embedding. The joint activation is then added to the self-activation, which should encourage the joint embedding to learn a residual function. The architecture is displayed in functional form as:

\begin{equation}
  \label{eq:u}
  \begin{gathered}
    state1 = \sigma ( W_1 s_1 + b_1 ) \\
    state2 = \sigma (W_2 s_2 + b_2) \\
    joint = [ state1; state2 ] \\
    residual = \max ( W_j joint + b_j , 0 ) \\
	embedding =  state1 + residual \\
    output = W_o embedding + b_o  \\
  \end{gathered}
\end{equation}

Where $s_1$ and $s_2$ represent the grid locations of agent 1 and agent 2, respectively. $W_1$, $W_2$, $W_j$, and $W_o$ are learned weight matrices and $b_1$, $b_2$, $b_j$, and $b_o$ are learned bias vectors. $\sigma$ represents the sigmoid operation, $[ ; ] $ represents vector concatenation, and $ + $ element-wise vector addition.

\subsection{TD-Targets}

We mathematically define the TD-Targets for the various settings.  We show functions only for agent 1 but agent 2's targets are the mirror image.

\begin{table}[H]
\centering
\caption{TD-Targets for I-DQN and J-DQN algorithms at timestep $j$. $a_1'$ represents agent 1's selected action, $\vec{a}'$ the joint action vector, $\theta_i$ represents network parameters for the $i^{th}$ agent, $NashEq$ converts q-values to payoffs and returns an action vector corresponding to a Nash equilibrium. }
\label{td-targets}
\begin{tabular}{|l|l|}
\hline
Algorithm & TD-Target For Agent 1  \\ \hline

IDQN & $ r_j + \gamma \max_{a_1'} Q(\phi_{j+1}, a_1'; \theta_1).$ \\
Set Controller & $ r_j + \gamma \max_{\vec{a}'} Q(\phi_{j+1}, \vec{a'}; \theta).$\\
Friend-Q & $ r_j + \gamma \max_{\vec{a}'} Q(\phi_{j+1}, \vec{a}'; \theta_1).$\\
Nash-Q & $ r_j + \gamma Q(\phi_{j+1}, NashEq(Q(\phi_{j+1},\vec{a}',1; \theta_1 ),  Q(\phi_{j+1},\vec{a}',2; \theta_2 )); \theta_1)$ \\ \hline

\end{tabular}

\end{table}

Set Controller and Friend-Q have the same function signature, but importantly, Friend-Q maintains a separate network for each agent and actions are chosen independently.

\section{Experimental Results} \label{results}

We show updates for convergence and average statistics over 1,000 episodes with random restart.

\subsection{Results on 11 x 11 Warehouse Environment} \label{warehouse-results}

We train and evaluate on the $ 11 \times 11 $ Warehouse Environment for 2 agents (Section \ref{sync-env}). The episode ends after the minimum of 100 steps and 4 boxes have been delivered.
\begin{table}[H]
\centering
\caption{Results for $11 \times 11$ Warehouse Environment with 4 boxes and reward of 10 per delivery. }
\label{warehouse-table}
\begin{tabular}{|l|l|l|l|l|}
\hline
Algorithm & Reward  (Mean $\pm$ Sd) & Steps & Q-Values & Convergence  \\ \hline
IDQN & 8.8 $\pm$ 4.9 & 61 & 7.9 & 2.7 M  \\
IDQN comp & 9.2 $\pm$ 4.3 & 52 & 9.6 & 2.1 M  \\
Set Controller & 9.6 $\pm$ 3.8 & 59 & 18.3 & 1.7 M \\
Friend & 0.82 $\pm$ 1.86 & 100 & 7.544 & N/A \\
Nash-Q & 0.11 $\pm$ 0.92 & 100 & 0.11 & N/A \\ \hline
\end{tabular}
\end{table}

\subsubsection{Analysis}
For this case, both competitive and cooperative IDQN converge. However, because of the larger grid size, set controller performs the best in terms of both convergence and rewards and Friend and Nash Q algorithms are not able to learn.

\subsection{Results for 5x5 Sync Environment  } \label{sync-results}

We train and evaluate on the $5 \times 5$ Sync Environment for 2 agents.  The Sync Environment is detailed in Section \ref{sync-env}.  Implicit coordination is a sine qua non for agents to receive a positive reward of 100. Table \ref{sync-table} shows average statistics and convergence steps for the suite of MADRL algorithms.

\begin{table}[H]
\centering
\caption{Results for 5x5 Sync Environment with reward of 100 for successful synchronization. }
\label{sync-table}
\begin{tabular}{|l|l|l|l|l|}
\hline
Algorithm & Reward  (Mean $\pm$ Sd) & Steps & Q-values & Convergence  \\ \hline
IDQN & 73.3 $\pm$ 23.3 & 5 & 28.7 & 0.35 M \\
Set Controller & 71.5 $\pm$ 19.5 & 7  & 85.0 & 0.35 M \\
Friend & 5.0 $\pm$ 20.6 & 21 & 14.8 & 0.35 M \\
Nash-Q & 0.29 $\pm$ 5.0 & 59 & -0.00 & N/A \\ \hline
\end{tabular}
\end{table}

\subsubsection{Analysis}

We observe that the IDQN and Set Controller setups converge well for both environments. Contrary to what we expected, IDQN converges to a slightly better reward. Yet, the Set Controller needs to learn the entire joint space for the 2 agents, so it needs more updates to give adequate coverage to each output node in the network. The Friend algorithm exhibits strong signs of learning as reflected in elevated Q-Values and nonzero rewards (for the tougher environment Sync). The Set Controller case is the gold standard for Friend since it explicitly optimizes what the Friend algorithm proxies with 2 separate Q-networks. We expect Friend's performance to continue to approach the gold standard with longer training times. Nash-Q fails to converge which we cover in more detail in Section \ref{nash-analysis}.

\subsection{Comparison of Asynchronous IDQN with Synchronous IDQN} \label{async-results}

We compare the asynchronous IDQN architecture with its synchronous counterpart. As explained in Section \ref{async-explanation}, the asynchronous setup maintains a single network and buffer to which both agents submit experience data, and from which both agents receive the latest weights once every 10,000 updates. We hoped this would speed up convergence by mitigating environment stochasticity and de-correlating experience data since both agents' experience is merged into a single buffer. We test using the $5 \times 5$ Sync environment and the IDQN variant as it maintains 2 networks and is stable.

\begin{table}[H]
\centering
\caption{Results for 5x5 Sync Environment With Asynchronous IDQN Variant. }
\label{async-table}
\begin{tabular}{|l|l|l|l|l|}
\hline
Algorithm & Reward  (Mean $\pm$ Sd) & Steps & Q-values & Convergence  \\ \hline
IDQN & 73.3 $\pm$ 23.3 & 5 & 28.7 & 0.35 M \\
IDQN-Asynchronous & 0.0 $\pm$ 0.1 & 100 & 0.0 & 0.35 M \\ \hline
\end{tabular}
\end{table}

\subsubsection{Analysis}

We were very surprised to discover that the asynchronous architecture did not converge.  We did not discover any blatant reasons for this, but our principal hypothesis is as follows. The network learns twice from a single game step, for which it receives mirrored data points. We expect this confuses the network and may lead to compounding errors. It is still surprising this issue exists with a large batch size and randomized memory replay, however.

\subsection{Comparison of Single Stream versus Residual Split Stream Architecture}

We compare the single stream with the residual split stream architecture, outlined in Section \ref{split-stream-explanation}.

\begin{table}[H]
\centering
\caption{Results on 5x5 Sync Environment for Split and Single Stream architectures. We test using the Set Controller setting which converges the quickest for both single and split stream. }
\label{split-stream-table}
\begin{tabular}{|l|l|l|l|l|}
\hline
Algorithm & Reward  (Mean $\pm$ Sd) & Steps & Q-values & Convergence  \\ \hline
Set Controller + Single Stream & 71.5 $\pm$ 19.5 & 7  & 85.0 & 0.35 M \\
Set Controller + Split Stream & 6.5 $\pm$ 22.8  & 6 & 5.7 & ~1 M \\ \hline
\end{tabular}
\end{table}

\subsubsection{Analysis}

The split stream architecture did not enhance training efficiency or stability in the manner we had hoped.  This is likely due to the simplified network architecture for split stream, for which we only learn a single-level joint embedding.  The single stream architecture conducts early fusion, and the joint representation passes through 3 dense layers; hence, it is able to learn more complex inter-agent dependencies. For the Sync environment on which we tested, coordination is the sine qua non for positive reward. Testing on a more forgiving environment, such as Warehouse, which does not demand such nuanced coordination, might allow the split stream architecture to outperform. The residual function concept is very compelling theoretically and has proven to work \cite{he2016deep}, so we do not think these results should be viewed as damning. Future work might explore keeping the single stream network in tact and simple introducing an auxiliary skip connection for the own agent's representation - thereby only enhancing, not detracting from, the model's expressive power.

\subsection{Comparison of Nash-Q Tie Breaking and No Nash Strategies}

Nash-Q theoretically converges only when there is a single optimal joint action at every step \cite{hu2003nash}. Yet, Hu et al. note that this is a restrictive, unrealistic precondition and as such, rules must be put in place to consistently break ties. We test on 4 permutations of tie-breaking rules, each defined in Section \ref{nash-explanation}.

\begin{table}[H]
\centering
\caption{Results for 5x5 Sync Environment For 4 Nash-Q Tie-Breaking Variants. }
\label{nashq-table}
\begin{tabular}{|l|l|l|l|l|}
\hline
Nash-Q Variant & Reward (Mean $\pm$ Sd) & Steps & Q-values & Convergence  \\ \hline
Max-Sum Tie-Break, Best-Sum-No-Nash & 0.29 $\pm$ 5.0 & 59 & -0.00 & N/A \\
Max-Sum Tie-Break, Greedy-No-Nash  & 0.85 $\pm$ 8.4 & 42 & 0.01 & N/A \\
Greedy-Tie-Break, Best-Sum-No-Nash & 1.1 $\pm$ 9.7  & 54 & 0.01 & N/A  \\
Greedy-Tie-Break, Greedy-No-Nash & 0.33 $\pm$ 5.7  & 73 & -0.01 & N/A \\ \hline
\end{tabular}
\end{table}

\subsubsection{Analysis} \label{nash-analysis}

Despite some variance in results, none of the permutations showed significant signs of reward or q-value convergence. Since there is no mathematical best practice regarding tie-breaking procedures for Nash equilibria, and its impact on convergence, it was worth exploring the gamut of available options. We were hoping that the socially optimal action profile choice: Max-Sum, would be a good proxy for the Set Controller. Yet, in the sync environment, if an agent is in the middle of the board, and attempting to reach the upper left corner, either an action of `Up' or action of `Left' will have the same effect. In these common instances of no optimal joint action, the model struggles to converge via implicit coordination. The case turns out to be similar for the other permutations as well. We learn from these experiments that the mere existence of multiple equilibria is problematic, and that even well-designed tie-breaking rules might not be sufficient to guarantee convergence. When no Nash equilibria are present, it is even more difficult to reason a stable strategy. Our approaches were grounded in the notion that the self-interested and cooperative incentive should align for fully cooperative settings. Yet the sensitivity of policy control over a 2-D payoff matrix proves extremely unstable regardless of network configurations.

\section{Discussion}

We show that a Nash equilibrium policy control can be applied to MADRL for multi-step MDPs under a key simplifying assumption.  While the Nash-Q algorithms did not show policy control stability, the Friend-Or-Foe simplification did. Over time, the Friend-Or-Foe's assumption of mirrored mutual best responses approaches, albeit slowly, a Set Controller, which explicitly selects actions for both agents. Concretely, for fully cooperative games, independent optimization of q-values over the joint action space can roughly approximate the explicit optimization of a single joint q-function given enough training time. Given the highly unstable nature of multi-agent decentralized coordination, and highly restrictive preconditions for theoretical convergence, we deem this an encouraging, albeit incomplete, accomplishment. The fact that Nash-Q did not converge is unsurprising. Hu et al. prove the convergence of Nash-Q only under "highly restrictive technical conditions", including but not limited to, the existence of a globally optimal action at every step \cite{hu2003nash}. Most environments, including ours, do not conform to this condition given that multiple actions can be optimal (with multiple equilibria). This is analogous to there not being a unique shortest path from one point to another in a grid without barriers.  We experiment with a multitude of stabilizing architectures, environments, and hyperparameter settings for Nash-Q.  Given that broad and deep experimentation did not give way to convergence lends credence to the difficulty of applying Nash-Q to multi-player, multi-stage games with nonsingular optimal joint actions.

We also experiment with a multitude of approaches for learning stabilization: dueling networks \cite{wang2015dueling}, double Q-networks \cite{van2016deep}, residual networks \cite{he2016deep}, priority experience replay \cite{schaul2015prioritized}, and asynchronous target fixing, to varying degrees of success. We demonstrate the theoretical advantages of each: double-Q - fair estimation of q-values, residual networks - coordination as residual function, prioritized experience replay - less sparse reward learning, and asynchronous target fixing - more environment stationarity from fixing weights.  We show suboptimal results for the residual split stream architecture and the asynchronous model but maintain that both are theoretically well-grounded.  We believe that the split stream architecture may need to learn a more robust joint embedding for complex coordination tasks and the asynchronous architecture should sample experience from a single agent to avoid generating multiple data points from a single game step.

\subsection{Conclusion}

We hope that our open-source code and 3 configurable multi-agent simulations will motivate future work at the fascinating intersection of Deep Q-Learning and Nash game theory. With simplifying assumptions (Friend algorithm), deep Nash models show signs of convergence and represent theoretically sound rule-based approaches for implicit coordination. Future work should focus on how to introduce stability in Nash-Q learning for multi-stage MDPs without a globally optimal action profile. Alternatively, properly identifying games phenotypes which necessitate Nash-Q Learning over simpler variants would be an appropriate first step toward a stronger theoretical grasp of when Nash-Q learning expedites coordination.

\newpage

\bibliography{bibliography}
\bibliographystyle{plain}

% \printbibliography

\section*{Appendix} \label{appendix}

\subsection{Hyperparameters}

\begin{table}[H]
\centering
\caption{Hyper parameters}
\label{my-label}
\begin{tabular}{|l|l|l|}
\hline
Hyper-parameter      & Warehouse Environment & Sync Environment  \\  \hline
Batch Size & 128 & 128 \\
Experience Replay Buffer size & 1,000,000 & 1,000,000 \\
Optimizer  & Adam \cite{kingma2014adam} & Adam \cite{kingma2014adam} \\
Learning Rate & 0.01   & 0.01 \\
Reward Constant & 10 & 100 \\
Learning Rate Decay & 1e-4 & 1e-4 \\
Loss & Mean Huber \cite{huber1964robust} & Mean Huber \cite{huber1964robust} \\
Initial Exploration $\epsilon$ & 1  & 1  \\
Final Exploration $\epsilon$  & 0.1 & 0.1 \\
Exploration Decay Factor  & 0.75 & 0.75 \\
Target Dissemination Freq (for asynchronous models) & 10,000 & 10,000 \\
Discount Factor   & 0.95 & 0.95   \\
Max Training Episode Length  & 100 & 100  \\ 
Max Test Episode Length & 100 & 100 \\
Number of Episodes for Evaluation & 2,000 & 2,000    \\
Experience Replay Burn-in  & 50,000 & 50,000      \\
Model Update Frequency  & 10 & 10  \\
Evaluation Frequency & 10,000 & 10,000 \\ \hline
\end{tabular}

\end{table}

\subsection{PredatorPrey Board - Instance}
\begin{figure}[ht]
    \centering
   \includegraphics[scale=0.065]{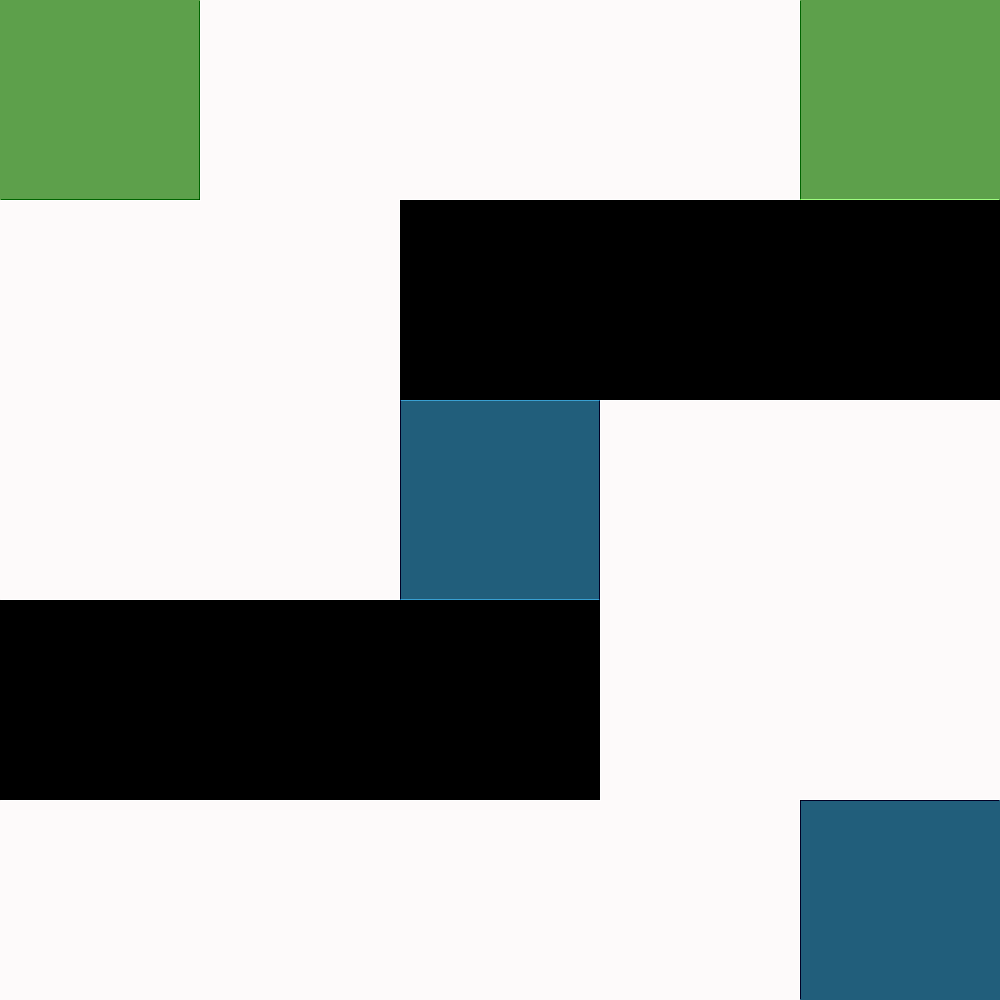}
\end{figure}
In the above image of 5x5 Predator Prey board, \textbf{Blue} represents Predators, \textbf{Green} represents Preys and barriers are represented by \textbf{Black} blocks.  Each predator receives this state representation (in the form of numpy grid) in addition to its own position.
Additionally, a video of the environment (of size 10x10) is rendered and can be viewed at \href{https://youtu.be/We0HvfFNd-E}{video link}. \\Similarly, an example of the warehouse environment can be seen at the given \href{https://www.youtube.com/watch?v=8ABmZDg9Ibk}{link}.

\subsection{Poster}

A poster board version of the paper can be seen here.

\begin{figure}[ht]
   \includegraphics[scale=0.2]{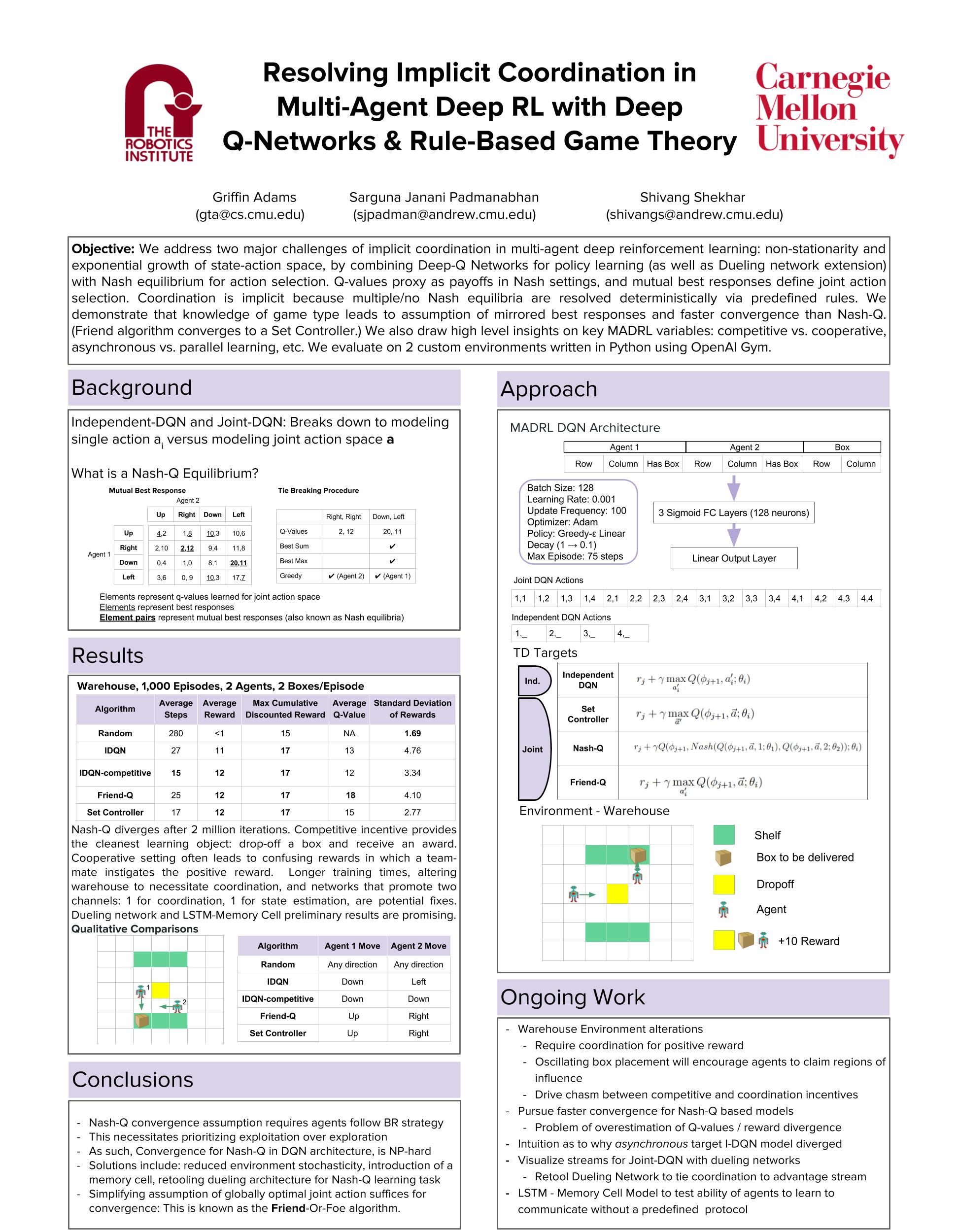}
\end{figure}

\end{document}